\documentclass{elsart}

\usepackage{graphics}
\usepackage{times}
\usepackage{amssymb}

\begin{document}

\begin{frontmatter}



\title{The STAR Silicon Strip Detector (SSD)}


\author[strasbourg]{L.~Arnold} 
\author[strasbourg]{ J.~Baudot} 
\author[strasbourg]{ D.~Bonnet} 
\author[nantes]{A.~Boucham} 
\author[nantes]{ S.~Bouvier} 
\author[nantes]{ J.~Castillo} 
\author[strasbourg]{ J.P.~Coffin} 
\author[nantes]{ C.~Drancourt} 
\author[nantes]{ B.~Erazmus} 
\author[nantes]{ L.~Gaudichet} 
\author[strasbourg]{ M.~Germain} 
\author[strasbourg]{ C.~Gojak} 
\author[warsaw]{J.~Grabski} 
\author[nantes]{ G.~Guilloux} 
\author[strasbourg]{ M.~Guedon} 
\author[strasbourg]{ B.~Hippolyte} 
\author[warsaw]{ M.~Janik} 
\author[warsaw]{ A.~Kisiel} 
\author[strasbourg]{ C.~Kuhn} 
\author[nantes]{ L.~Lakehal-Ayat} 
\author[nantes]{ F.~Lefevre} 
\author[nantes]{ C.~Le~Moal} 
\author[warsaw]{ P.~Leszczynski} 
\author[strasbourg]{ J.R.~Lutz} 
\author[warsaw]{ A.~Maliszewski} 
\author[nantes]{ L.~Martin} 
\author[nantes]{ T.~Milletto}
\author[warsaw]{ T.~Pawlak} 
\author[warsaw]{ W.~Peryt} 
\author[warsaw]{ J.~Pluta} 
\author[warsaw]{ M.~Przewlocki} 
\author[warsaw]{ S.~Radomski} 
\author[nantes]{ O.~Ravel} 
\author[nantes]{ C.~Renard} 
\author[nantes]{ G.~Renault} 
\author[nantes]{ L.M.~Rigalleau} 
\author[nantes]{ C.~Roy} 
\author[nantes]{ D.~Roy} 
\author[strasbourg]{ C.~Suire} 
\author[warsaw]{ P.~Szarwas} 
\author[strasbourg]{ A.~Tarchini}

\address[strasbourg]{Institut de Recherche Subatomique, 23 rue du Loess, 67037 Strasbourg, France}
\address[nantes]{Subatech, 4 rue A.~Kastler BP 20722, 44307 Nantes, France}
\address[warsaw]{Warsaw University of Technology, Koszykowa 75, 00-662 Warsaw, Poland}

\begin{abstract}
  The STAR Silicon Strip Detector  (SSD) completes the three layers of
  the Silicon  Vertex Tracker (SVT)  to make an inner  tracking system
  located inside  the Time Projection Chamber  (TPC).  This additional
  fourth layer  provides two dimensional hit position  and energy loss
  measurements for  charged particles, improving  the extrapolation of
  TPC  tracks through  SVT hits.   To match  the high  multiplicity of
  central  Au+Au collisions  at RHIC  the double  sided  silicon strip
  technology was  chosen which makes  the SSD a half  million channels
  detector. Dedicated electronics have  been designed for both readout
  and control. Also  a novel technique of bonding,  the Tape Automated
  Bonding (TAB), was used to fullfill the large number of bounds to be
  done. All  aspects of  the SSD are  shortly described here  and test
  performances  of produced  detection  modules as  well as  simulated
  results on hit reconstruction are given.
\end{abstract}

%\begin{keyword}

%\PACS 
%\end{keyword}
\end{frontmatter}

% main text
\section{Introduction}
%\label{}

The  STAR Silicon  Strip  Detector (SSD)  \cite{SSD_TechnicalProposal}
constitutes the  fourth layer of the inner  tracking system. Installed
between  the Silicon  Vertex  Tracker (SVT)  and  the Time  Projection
Chamber (TPC), the  SSD will enhance the tracking  capabilities of the
STAR  experiment  by  measuring  accurately the  two  dimensional  hit
position and energy loss of charged particles. It aims specifically at
improving  the  extrapolation  of  TPC  tracks through  SVT  hits  and
increasing  the  average number  of  space  points  measured near  the
collision  thus  increasing  the  detection efficiency  of  long-lived
meta-stable particles.
 
The SSD is placed at a distance of 230 mm from the beam axis, covering
a pseudo-rapidity range of $|\eta|<1.2$ which leads to a total silicon
surface close to 1 $m^2$.

\begin{figure}
{\par\centering \resizebox*{0.9\textwidth}{!}{\rotatebox{0}{\includegraphics{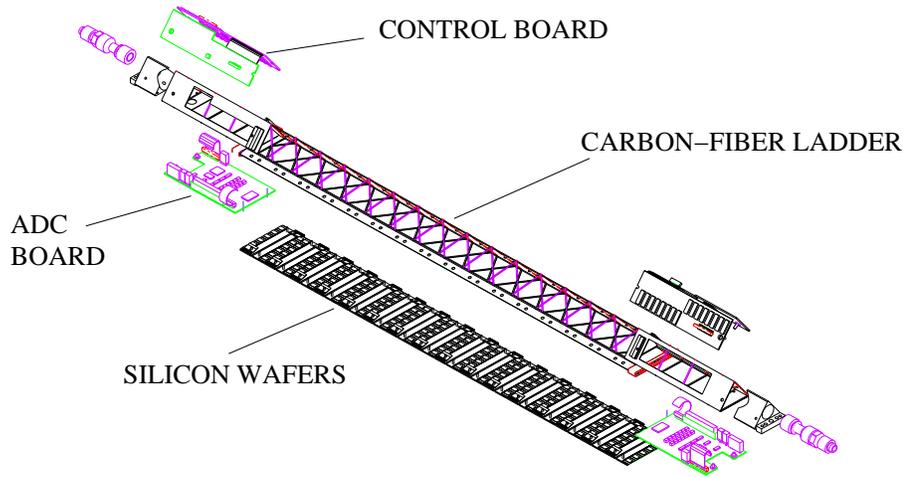}}} \par}
\caption{An SSD ladder showing separately its components.\label{SSD_Ladder} }
\end{figure}

The design of  the SSD is based on two  clamshells, each containing 10
carbon-fiber ladders.  Each  ladder (figure \ref{SSD_Ladder}) supports
16 wafers using double-sided  silicon strip technology (768 strips per
side) and connected  to the front-end electronics (6  ALICE 128C chips
per  side) by  means of  the  Tape Automated  Bonded (TAB)  technology
\cite{SSD_TAB_Stephane}. The ladders are  tilted with respect to their
long axis,  allowing the  overlap of the  detectors in  the transverse
plane for better hermiticity  and alignment performances.  A bus cable
transports  the analog signals  along the  ladder to  two 10  bits ADC
boards  installed at both  ends. After  digitization, the  signals are
sent  to Readout Boards  which are  linked to  the DAQ  system through
Giga-link optics  fibers. The whole  system is remotely  controlled to
monitor  powers and  temperature but  also to  calibrate and  tune the
front-end  electronics. The  cooling system  is based  on  an air-flow
through  the ladder  which is  embedded in  a mylar  sheet.  The total
radiation length has been estimated to be around $1\%$.

\section{Mechanics}
The mechanical design  is constrained by : 1) the  inner radius of the
TPC (including a  safety margin in order to  avoid high voltage breaks
from the  TPC field  cage); 2) limited  material budget;  3) stability
with  respect  to  temperature  variations and  minimizatiuon  of  the
sagging  due  the  weight of  the  modules  in  order to  insure  good
alignment  precision.  Each  ladder  is  1060  mm  long,  featuring  a
triangular  cross-section of  40  mm  base. The  ladders  are made  of
high-modulus carbon-fiber which allows for  both a good rigidity and a
low material budget. The saggita of a loaded ladder has been estimated
by simulation to  be lower than 100 $\mu$m.  The  SSD modules are glued
on the carbon  ladder by an elastic epoxy glue  in order to accomodate
the different  thermal expansion coefficients  of silicon and  carbon. 
Two sets of ten ladders are assembled on C-shaped ribs which allow for
installation on the common SVT/SSD support structure.

\section{Wafers}
\begin{figure}
  {\par\centering
    \resizebox*{0.6\textwidth}{!}{\rotatebox{0}{\includegraphics{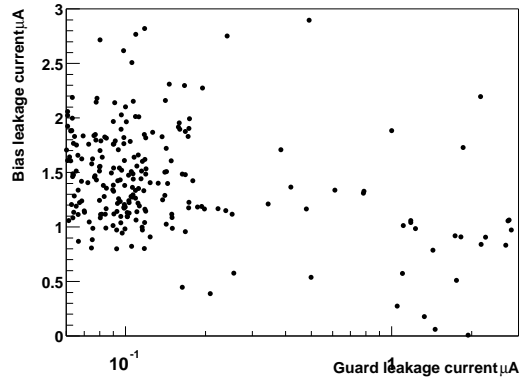}}}
    \par}
\caption{Distribution of the leakage current of the bias ring with respect to the leakage current of the guard ring of the wafers. It demonstrates that the leakage current through the guard ring due to cut-off and surface effects on the silicon sensor stay well below or do not affect the bias leakage current which in turn is reasonnably small in average and usually lower than $3 \mu$A.\label{SSD_waferIbVsIg} }
\end{figure}

\begin{figure}
  {\par\centering
    \resizebox*{0.6\textwidth}{!}{\rotatebox{0}{\includegraphics{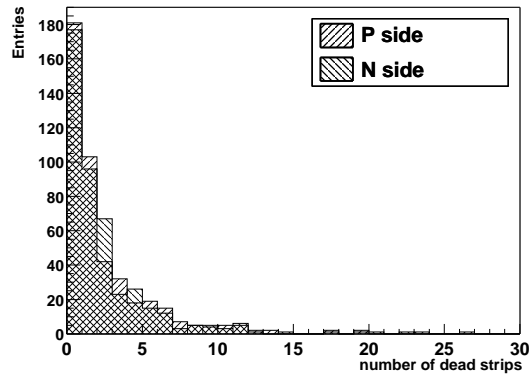}}}
    \par}
\caption{Distribution of the number of dead strips identified per side of the wafers.\label{SSD_waferDeads} }
\end{figure}

The design of the silicon wafers was inspired by the initial design of
the ALICE  experiment silicon strip  detector \cite{SSD_ALICE_TP}. The
wafers are 75 mm  by 42 mm sized and 300 $\mu$m  thick. The strips are
oriented perpendicularly with respect  to the longest dimension of the
detector on both  sides (double-sided technology), with a  pitch of 95
$\mu$m (768 strips  per side). The insulation between  strips on the N
side  is   achieved  by  p-spraying,  and  the   biasing  scheme  uses
punch-through  technology. The bias  ring surrounded  by a  guard ring
define an  active area of 73  mm by 40 mm.  The strips on the  P and N
sides  are  inclined  by  35  mrad  with respect  to  each  other  and
symetrically with respect to the  wafer edge. Combined with the charge
correlation  due to  the  double-sided technology,  this stereo  angle
allows an efficient 2-dimension  impact point reconstruction in a high
multiplicity    environment    with     a    low    ambiguity    yield
\cite{SSD_MultipSandra}.  The  pitch is chosen  in order to  achieve a
minimum resolution (digital resolution) of 30 $\mu$m in the transverse
plane and $800\mu$m along the  Z axis.  During production, all the 440
detectors were  fully characterized with a  probe-station allowing the
measurement  of their  depletion voltage  and leakage  current (figure
\ref{SSD_waferIbVsIg}).  The  detectors   can  currently  be  operated
between 20 V and 50 V.  Moreover, the corresponding 680000 strips have
been scanned to identify  every defective strips \cite{SSD_Test_Chip}. 
A  very low average  number of  dead strips  have been  measured, well
below   $1\%$,  see   figre   \ref{SSD_waferDeads}.  Prototypes   were
irradiated in order to test their capability to cope with the expected
fluxes \cite{SSD_Irradiation}.

\section{Front-end electronics and modules}
\begin{figure}
{\par\centering \resizebox*{0.5\textwidth}{!}{
\includegraphics{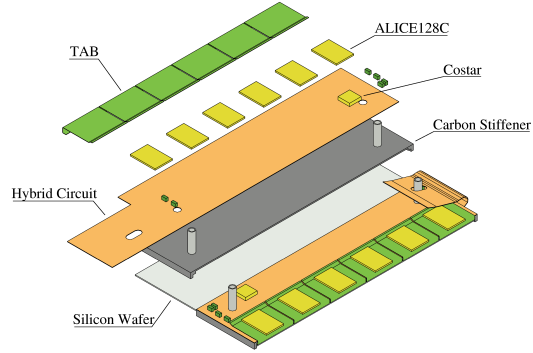}
} \par}
\caption{Schematic view of an SSD module with its compnents.\label{SSD_Module} }
\end{figure}

\begin{figure}
{\par\centering \resizebox*{0.8\textwidth}{!}{\rotatebox{0}{\includegraphics{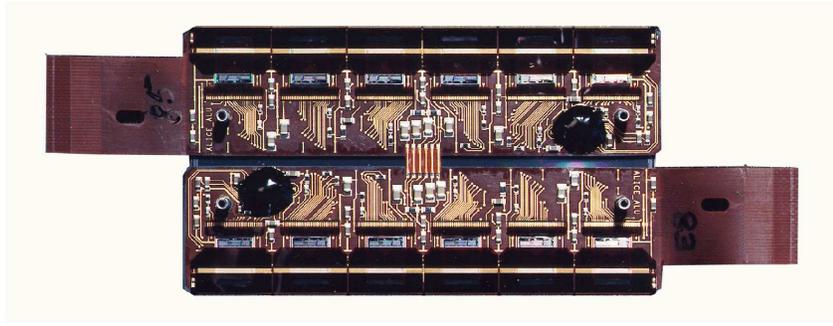}}} \par}
\caption{Photo of an assembled module viewed from the hybrid side.\label{SSD_ModuleTabed} }
\end{figure}

Each  of  the  320  detection  modules  (figure  \ref{SSD_Module})  is
composed  of  a  silicon  wafer,two  hybrid circuits  (one  per  side)
supporting  the  front-end  electronics  and  a  mechanical  structure
\cite{SSD_Design}.   The front-end electronics  is constituted  by six
ALICE 128C chips per side  \cite{SSD_ALICE_Chip}.  The ALICE 128C is a
low  power chip,  specially designed  for the  STAR and  ALICE silicon
strip detectors using the AMS 1.2 $\mu$m CMOS technology.  Each of the
128  analog input  channels  of the  chips  allows the  amplification,
shaping and  storage of  the charge collected  on a single  strip. The
shaping time is adjustable between 1.2 and 2 $\mu$s, the dynamic range
extends up  to $\pm$13 MIPs  ( signals collected  on each side  of the
wafer have  different polarity) with  a nominal noise  of 290 $e$  + 8
$e/pF$. An analog mutiplexer allows the sequential readout of the data
at a rate up to 10 MHz.The chip is controlled through JTAG protocol in
order  to tune and  monitor remotely  its performance.   The front-end
electronics is supported by a hybrid circuit which is a copper printed
kapton tape glued  on a carbon stiffener.  The  main components on the
hybrids  are  ALICE128C  chips   and  a  COSTAR  (COntrol  STAR)  chip
\cite{SSD_COSTAR} designed for slow control purposes : temperature and
leakage current  measurements as well as  pedestal compensation.  Tape
Automated Bonding (TAB) is used  for the connection between the strips
on  the  detector  and the  analog  inputs  of  the ALICE  128C  chips
\cite{SSD_TAB_Christophe}.   This bumpless  technology is  based  on a
kapton microcable on which are printed copper strips,
%(figure \ref{SSD_Tab}) 
the flexibility  of the cable allowing  the folding of  the hybrids on
top  of  the detector  in  order to  make  the  detection module  very
compact,  see figure  \ref{SSD_ModuleTabed}.  Furthermore, this  cable
plays the role of a pitch adaptator between the 95 $\mu$m pitch of the
detector  and the  $44 \mu$m  pitch of  chips on  the hybrid.  The TAB
technology is  also used for the  connection between the  chip and the
hybrid.  During the production stage, the chip is first connected to a
TAB  tape  and  then  fully  tested  \cite{SSD_Test_Chip}  before  its
connection to the detector.  An internal pulser permits testing of all
channels individually both during stages  of assembly and on the final
installation in  STAR.  After the  assembly of modules, each  strip is
calibrated with a dedicated laser test bench.

%\begin{figure}
%{\par\centering \resizebox*{0.4\textwidth}{!}{\rotatebox{0}{\includegraphics{SSD_Tab.eps}}} \par}
%\caption{Photo of a ALICE 128C chip bonded on its TAB cable. The copper strips fan-out to the detector are visible on the tape which is still in its support frame used for test.\label{SSD_Tab} }
%\end{figure}

\section{Readout electronics}
The Control Board which receives the signals from the modules and also
takes care  of JTAG communication and latchup  protection is connected
to the hybrid  with a kapton cable. The  analog signals are transfered
from the  Control Board to the ADC  Board, both located at  the end of
the ladder  (one set  per side  and per ladder).  The ADC  board reads
sequentially the data  of a full ladder at 10 MHz  and digitizes to 10
bits. The  analog signal  (50 mV/MiP) spans  a range  of 1 V.  The two
piece Control Board is joined with  a flex cable and folded to conform
to the triangular carbon ladder. The Control Board is connected to the
ADC Board  which is installed at  the bottom edge of  the ladder.  The
interface between  the detector (ADC,  JTAG control of the  chips) and
the  STAR system  (DAQ, Slow-control,  trigger) is  done  through four
Readout Boards located  on the TPC wheels. All  the functionalities of
the  Readout Board  are performed  by  an FPGA  programable chip.  The
system is able to read all the channels in less than 5 ms.

\section{Performances}
\begin{figure}
{\par\centering \resizebox*{0.7\textwidth}{!}{\includegraphics{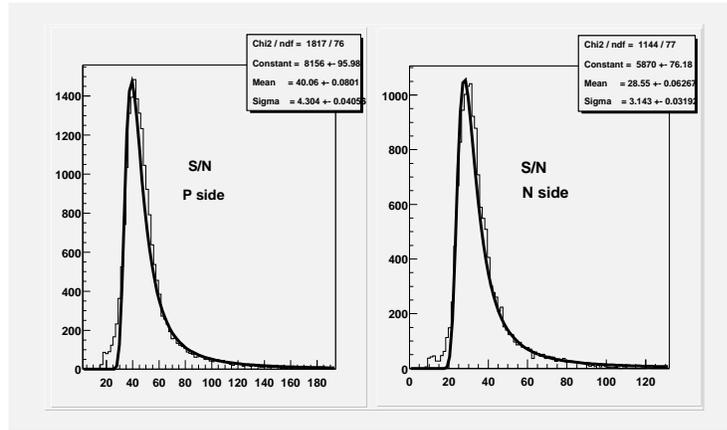}} \par}
\caption{Typical signal-to-noise distributions obtained with SSD prototypes.\label{SSD_SignalToNoise} }
\end{figure}

\begin{figure}
  {\par\centering
    \resizebox*{0.7\textwidth}{!}{\includegraphics{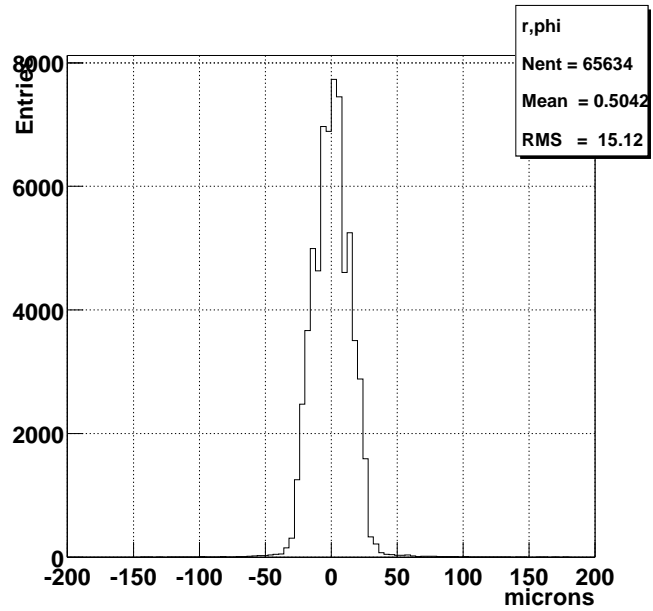}}
    \par}
\caption{Distribution of the error on the position of the hit obtained with a detailed simulation of the SSD ($r/\phi$ direction).\label{SSD_PositionResolution} }
\end{figure}

The most  important characteristics of the SSD  are spatial resolution
and  detection efficiency.   The first  is  driven by  the pitch,  the
diffusion of  the charge cloud  in the the  wafer bulk, the  signal to
noise ratios and alignment  accuracy.  The efficiency is determined by
the fraction of  dead strips on the detector and  dead channels on the
ALICE  128C  chips, the  performance  of  the  connection between  the
detector  and the front  end electronics  and signal-to-noise  ratios. 
Prototypes of the detection modules, as well as of the Control and ADC
boards   were   tested  off-line   and   in-beam   at  CERN.    Figure
\ref{SSD_SignalToNoise}  shows typical  histograms  of signal-to-noise
ratios  obtained from in-beam  tests. The  2-D spatial  resolution was
estimated from  the beam-test data by reconstructing  impact points on
both sides  of detectors and  combining these information.  The values
are   around  15   $\mu$m   in  the   $r/\phi$   direction  (   figure
\ref{SSD_PositionResolution})and 750 $\mu$m in the direction along the
strips      (corresponding      to      the     beam      axis      in
STAR)\cite{SSD_Performances}. These spatial  resolutions are driven by
the  intrinsic resolution  on each  side,  which are  typically of  20
$\mu$m  and the  stereo  angle. Due  to  the quality  of the  selected
detectors, chips and to the TAB  connection, one can expect a level of
dead channels below $2\%$ in  average. This quality is achived through
full tests of  all the components at all stages  of the production and
assembly phases  : detector, chip, hybrid, module,  ladder, clamshell. 
All  information   collected  during  these  tests  is   stored  in  a
database\footnote{wwwstar-sbg.in2p3.fr} which  may be accessed  at any
time  and  from   anywhere\cite{SSD_DataBase}.  This  database  system
provides information for monitoring the production, selecting elements
for  the assembly  and eventually  checking  the status  of the  whole
detector.

\begin{figure}
  {\par\centering
    \resizebox*{0.7\textwidth}{!}{\includegraphics{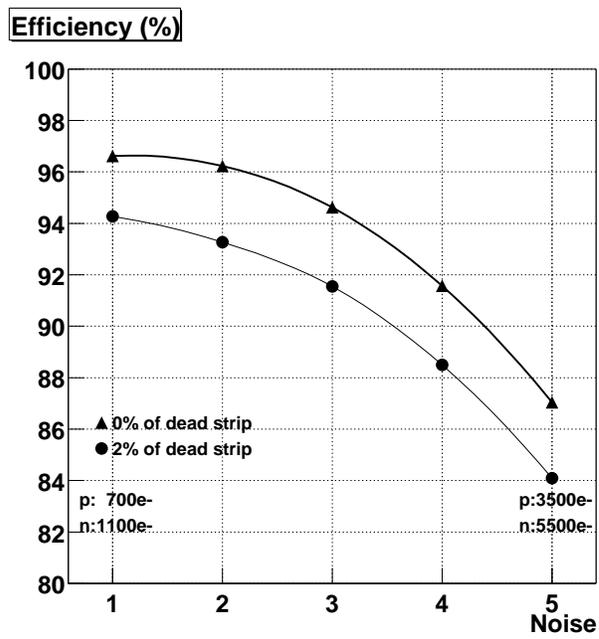}}
    \par}
\caption{Dependence of 2-D hit reconstruction efficiency on noise level and dead channels fraction.\label{SSD_EfficiencyVsNoise} }
\end{figure}

A detailed simulation of the  detector as well as a hit reconstruction
algorithm have been developped  to evaluate the global performances of
the SSD  \cite{SSD_MultipBoris}. The method  uses the matching  of the
signal amplitudes on  both side of the detector  for the same particle
to  solve ambiguous situation  where several  particle hits  are close
together. In nominal  conditions for the noise and  the number of dead
channels the  hit reconstruction  reaches 95~\% efficiency  with 98~\%
purity. The  dependence of the efficiency  on the noise  level and the
number     of     dead     strips     is    pictured     in     figure
\ref{SSD_EfficiencyVsNoise}. The general  trend is that the efficiency
decreases steadily with  the average number of dead  strips whereas it
is much  more stable  with respect  to the noise.  The purity  is less
sensitive to  this condition  and stays above  $98\%$ even  for $10\%$
dead strips or 4 times nominal noise level.

\section{Summary}

The Silicon Strip  Detector for STAR with around  1~m$^2$ of sensitive
surface and  half a million channels  is one of  the largest one-layer
system using  the doubled sided silicon strip  technology, see summary
table \ref{SSD_summary}. A compact module design was achieved by using
a novel  Tape Automated  Bonding method to  connect both sides  of the
silicon  wafers  to  the  front  end  electronics.   Other  innovative
solutions were used  like air flow cooling and  full remote control of
the  front-end electronics  to  minimize the  material  budget and  to
handle the large amount of parameters needed to operate the SSD.

Position resolutions of $15 \mu$m  in the transverse to beam direction
and $750 \mu$m in the beam direction were measured with beam test.
%In-beam tests of final design detection modules demonstrated that the resolutions of $15 \mu$m in the transverse to beam direction and $750 \mu$m in the beam direction was indeed achievable. 
A full simulation with realistic noise and dead channel conditions was
performed  and yelded  a 95~\%  efficiency with  98~\% purity  for hit
reconstruction.

The detector is currently in production phase and will be installed in
STAR  in 2002.  Exhaustive testing  of  all the  sensitive pieces  are
conducted and results are stored  in a database for further use during
the operation of the detector.

\begin{table}
\begin{center}
\begin{tabular}{||c|c||}
\hline
\hline
\multicolumn{2}{||c||}{\bf General layout} \\
\hline
\hline
radius                    & 230~mm \\
\hline
ladder length             & 1060~mm \\
\hline
acceptance                & $|\eta|<1.2$ \\
\hline
$\#$ of ladders           & 20 \\
\hline
$\#$ of wafers per ladder & 16 \\
\hline
total $\#$ of wafers      & 320 \\
\hline
\hline
\multicolumn{2}{||c||}{\bf Silicon wafers characteristics} \\
\hline
\hline
$\#$ of sides per wafer      & 2 \\
\hline
$\#$ of strips per side      & 768 \\
\hline
total readout channels       & 491520 \\
\hline
silicon wafer sensitive area & 73~x~40~mm \\
\hline
total silicon surface        & 0.98~m$^2$ \\
\hline
wafer pitch                  & 95~$\mu$m \\
\hline
$r\phi$ resolution           & $20\> \mu m$ \\
\hline
z resolution                 & $740\> \mu m$ \\
\hline
operating voltage            & $20 - 50$ V \\
\hline
leakage current for one wafer & $1 - 2\> \mu A$ \\
\hline
\hline
\multicolumn{2}{||c||}{\bf Readout front-end electronics} \\
\hline
\hline
$\#$ of input channels per circuits & 128 \\
\hline
total $\#$ of circuits              & 3840 \\
\hline
dynamical range                     & $\pm 13$ MIPs \\
\hline
shaping time                        & $1.2 - 2 \mu s$ \\
\hline
Signal / Noise                      & $30 - 50$ \\
\hline
SSD total readout time              & $<$~5~ms \\
\hline
\hline
\multicolumn{2}{||c||}{\bf Expected performances} \\
\hline
\hline
dead channels level           & $\sim 2\%$ \\
\hline
hit reconstruction efficiency & $\sim 95\%$ \\
\hline
hit reconstruction purity     & $\sim 98\%$ \\
\hline
\hline
\end{tabular}
\caption{Summary of the SSD characteristics and performances.\label{SSD_summary} }
\end{center}
\end{table}


\begin{thebibliography}{00}


% notes:

% \bibitem{label2}
% If there is a note, it should come last:
% \bibitem{label3} \note
% \end{subbibitems}

\bibitem{SSD_TechnicalProposal} STAR SSD Collaboration, Proposal for a silicon strip detector for STAR, STAR note SN-0400

\bibitem{SSD_TAB_Stephane} S. Bouvier, TAB : a packaging technology used for silicon strip detector to front end electronics interconnection, Proceedings of the $4^{th}$ workshop on electronics for LHC experiments, Rome, Sept. 1998.

\bibitem{SSD_Design} J.R. Lutz et al., Detector and front-end electronics for ALICE and STAR silicon strip layers, Proceedings of the $4^{th}$ workshop on electronics for LHC experiments, Rome, Sept. 1998

\bibitem{SSD_ALICE_TP} ALICE Collaboration, Technical Proposal, CERN/LHCC/95-71.

\bibitem{SSD_MultipSandra} S. Giliberto et al., Performances of double-sided silicon strip detectors in a high multiplicity environment of the ALICE experiment at LHC., ALICE note INT-99-53

\bibitem{SSD_MultipBoris} B. Hyppolyte et al., Silicon strip detector reconstruction chain for the STAR experiment, STAR note SN-0427                 

\bibitem{SSD_Irradiation} M. Germain et al., Irradiation of a silicon-strip detector and readout chips for the ALICE experiment at LHC, NIM A434 (1999) 345-357\\
M. Germain et al., Irradiation of silicon-strip detector for the ALICE experiment at LHC, ALICE note INT-01-02

\bibitem{SSD_ALICE_Chip} L. Hebrard et al., ALICE 128C : a CMOS full custom ASIC for the readout of silicon strip detectors in the ALICE experiment, proceeding of the $3^{rd}$ workshop on electronics for LHC experiments, London, Sept. 1997

\bibitem{SSD_COSTAR} D. Bonnet et al., Control System of the silicon microstrip layer of the STAR experiment, Proceedings of the ICALEPS'99 conference, Trieste, Oct. 99

\bibitem{SSD_Test_Chip} J.R. Lutz et al., Production tests of microstrip detector and electronic front-end modules for the STAR and ALICE trackers, Proceedings of the $5^{th}$ workshop on electronics for LHC experiments, Krakow, Sept. 2000

\bibitem{SSD_Performances} L. Arnold et al., Beam-test of double-sided silicon strip detector, ALICE note INT-98-05\\
C. Suire et al., Off and In-beam tests of silicon strip detectors for the ALICE experiment at LHC, ALICE note INT-99-22\\
F. Retiere et al., Performances of double-sided silicon strip for the ALICE experiment at LHC, ALICE note INT-99-36

\bibitem{SSD_TAB_Christophe} C. Suire et al., TAB connection for the silicon strip detector in STAR, STAR note SN-0431

\bibitem{SSD_DataBase} J.~Baudot, M.~Janik, W.~Peryt, P.~Szarwas et al., A database system for the production of the silicon strip detector for STAR, Proceedings of the $8^{th}$ International Conference on Accelerator and Large Experimental Physics Control Systems, San Jose, 27-30 November

\end{thebibliography}
\end{document}